\documentclass[aip,reprint,amsmath,amssymb,onecolumn]{revtex4-1}
\usepackage{graphicx,subfigure}
\usepackage{multirow}
\usepackage{chemfig,chemarrow}
\usepackage{tabularx,makecell}
\usepackage{caption,color,booktabs,float}

\newcommand{\cb}{\boldsymbol c}

\newcommand{\Pb}{\boldsymbol P}
\newcommand{\pb}{\boldsymbol p}
\newcommand{\xb}{\boldsymbol x}

\newcommand{\phib}{\boldsymbol \phi}
\newcommand{\psib}{\boldsymbol \psi}
\newcommand{\thetab}{\boldsymbol \theta}
\newcommand{\Omegab}{\boldsymbol \Omega}

\newcommand{\Ab}{\boldsymbol A}

\newcommand{\Eb}{\mathbb E}
\newcommand{\Qb}{\boldsymbol Q}
\newcommand{\varphib}{\boldsymbol \varphi}
\newcommand{\Prob}{\text{Prob}}

\newcommand{\com}[1]{{  #1}}

\begin{document}

\title{Two-scale large deviations for chemical reaction kinetics through second quantization path integral}

\author{Tiejun Li}\email{tieli@pku.edu.cn}
\affiliation{Laboratory of Mathematics and Applied Mathematics and  School of Mathematical Sciences, Peking University, Beijing 100871, P.~R. China}
\author{Feng Lin}\email{math\_linfeng@pku.edu.cn}
\affiliation{Laboratory of Mathematics and Applied Mathematics and  School of Mathematical Sciences, Peking University, Beijing 100871, P.~R. China}


\date{\today}

\begin{abstract}
Motivated by the study of rare events for a typical genetic switching model in systems biology,
in this paper we aim to establish the general two-scale large deviations for chemical reaction systems. We build a formal approach to explicitly obtain the large deviation rate functionals for the considered two-scale processes based upon the second-quantization path integral technique. We get three important types of large deviation results when the underlying two times scales are in three different regimes. This is realized by singular perturbation analysis to the rate functionals obtained by path integral. We find that the three regimes possess the same deterministic mean-field limit but completely different chemical Langevin approximations. The obtained results are natural extensions of the classical large volume limit for chemical reactions.\com{ We also discuss its implication on the single-molecule Michaelis-Menten kinetics.} Our framework and results can be applied to understand general multi-scale systems including diffusion processes.
\end{abstract}
\maketitle

\section{Introduction}

In recent years there has been a growing interest in studying the rare transitions for fast-slow stochastic dynamics in biology [\onlinecite{Assaf, Lv14, Lv15, LiLin, Ge, Newby10, Newby13, Wang, Wolyness}].  In computational neuroscience, the stochastic hybrid system is utilized to model the fast switching of ion channels and the membrane voltage evolves according to a dynamics which depends on the ion channel states.  In systems biology, people are interested in the phenotypic switching of the cells modeled by the central dogma, which involves fast switching of DNA states between active and inactive states and the transcriptional and translational processes with different rates depending on the DNA states. In both cases, the transition rates and the most probable transition paths between different stable fixed points  are issues being investigated in the literature. The main approaches include the WKB asymptotics and the path integral formulations. However, mathematically it falls in the field of large deviation theory (LDT) \cite{Varadhan, Freidlin, Veretennikov2000, Veretennikov1999, Liptser1996, Shwartz1995, Hugo} and the rigorous results for these types of problems are very limited \cite{LiLin}. It is also meaningful to remark that there is a close connection between the LDT and the popular landscape theory for biological systems\cite{Lv14,Lv15,ZhouLi}.

\begin{figure}
\centering
\includegraphics[width=0.45\textwidth]{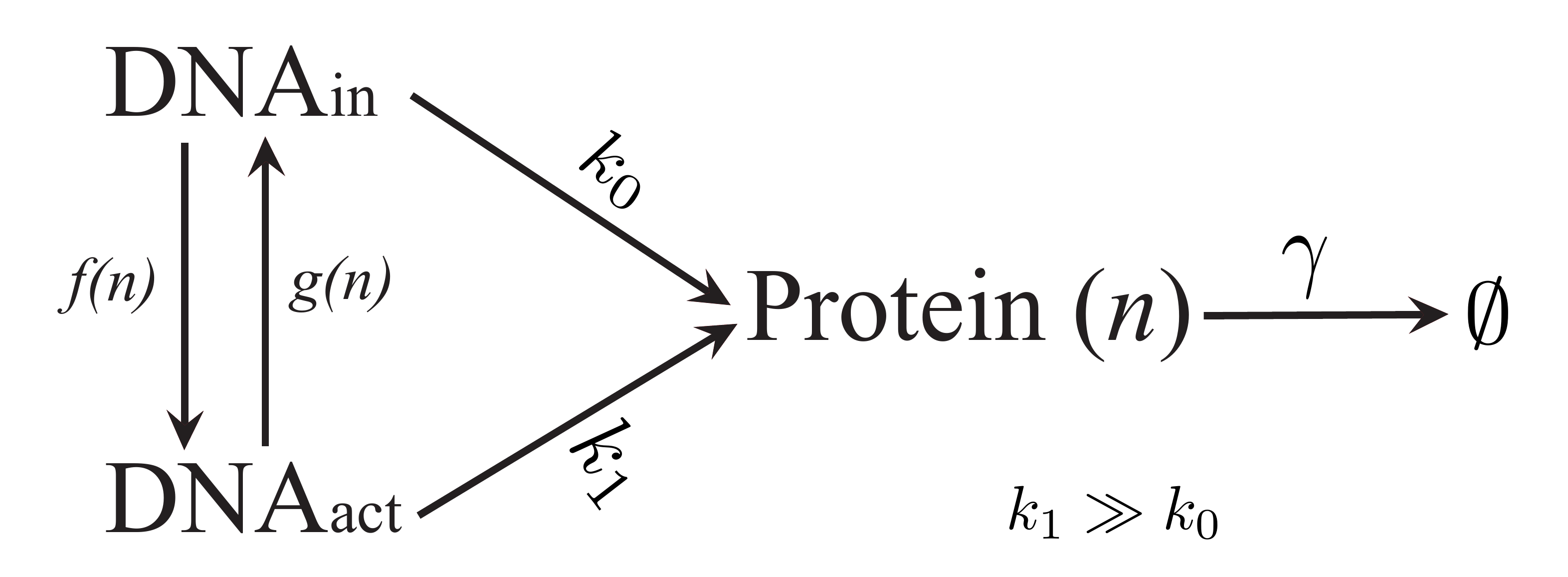}
\hskip 0.02\textwidth\includegraphics[width=0.45\textwidth]{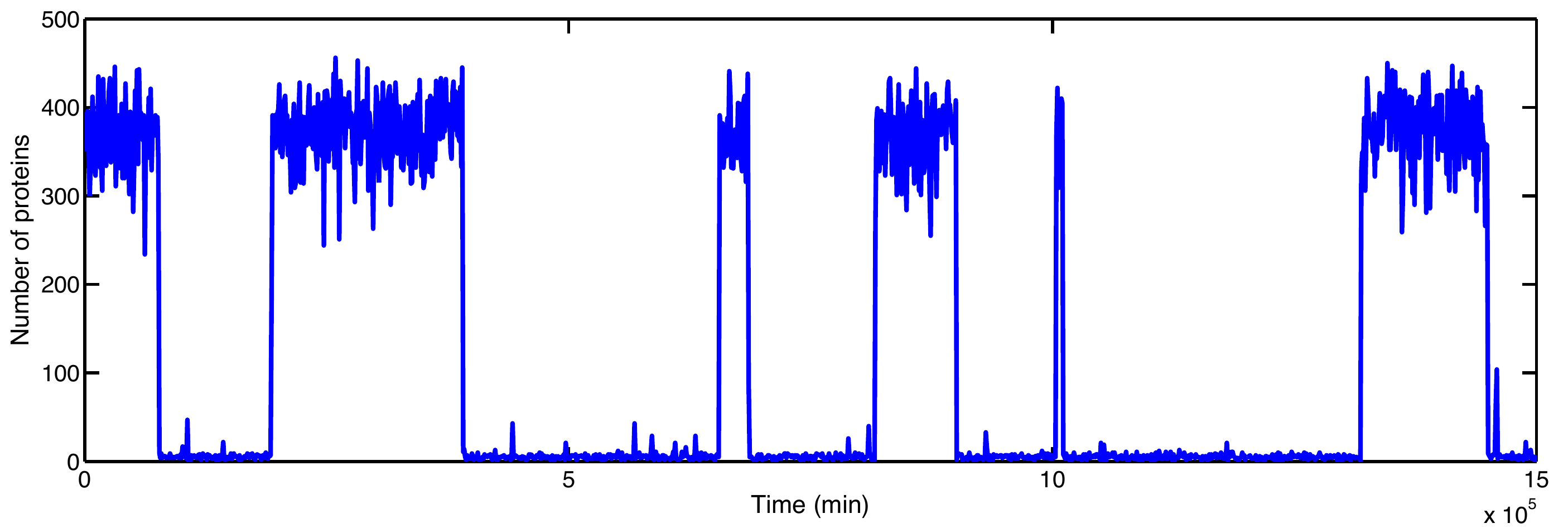}
\caption{A typical fast-slow genetic switching model considered in systems biology. Left panel: Schematics of the chemical reaction schemes, where the switching rates $f(n),~g(n)$ are usually large. Right panel: Direct Monte Carlo simulations of the genetic switching model. The bi-stability is clearly observed from the time series of protein copy numbers.}\label{fig1}
\end{figure}

In this paper, we will continue our program to study the two-scale large deviations for chemical kinetic systems. To illustrate our points more concretely, let us consider a canonical genetic switching model \cite{Ge,Wang} in systems biology as shown in Fig. \ref{fig1}. Dynamics of this self-regulating genetic system can be described by the following chemical master equation
\begin{eqnarray}\label{eq:Intro1}
{\partial_t}\Pb(n,t)&=&\left(
\begin{array}{cc}
k_0 & 0  \\
 0 & k_1
\end{array}
\right) (\Eb_{n}^{-1}-1)\Pb(n,t)+\gamma (\Eb_{n}^{1}-1)[n\Pb(n,t)]+
\left( {\begin{array}{cc}
   { - g(n)} & {f(n)}  \\
   {g(n)} & { - f(n)}  \\
\end{array}} \right)
\Pb(n,t),
\end{eqnarray}
where the two-component vector $\Pb(n,t)=(P_0(n,t),P_1(n,t))^T$, and $P_j(n,t)$ is the probability distribution function that the system has $n$ protein copy numbers in the DNA active $(j=1)$ or inactive state $(j=0)$. The raising or lowering operator $\Eb_{n}^{k}$ is defined through $\Eb_{n}^{k} h(n)=h(n+k)$ for any function $h$ depending on $n$.  $k_0$ and $k_1$ are protein synthesis rates, $\gamma$ is the degradation rate constant, and $f(n),~g(n)$ are switching rates between two DNA states.

The biologically relevant parameter setup is $\gamma \sim O(1)$  and $k_0/\gamma,~k_1/\gamma$ both large. We will not consider more detailed regimes concerning the magnitudes of $k_{1}$ and $k_0$  although one usually has $k_{1}\gg k_0$ in realistic situations. This does not affect the main point in this paper. In this case, the average number of proteins at steady state is of order $k_1/\gamma$. Now let us define the small parameter $\epsilon\approx \gamma/k_1$ or $\gamma/k_0$, thus the characteristic number of proteins is $n\sim O(\epsilon^{-1})$. In our fast-slow genetic switching model, we define the switching rates $f(n),g(n)\sim O(\epsilon^{-\alpha})$, and the realistic situations can be classified into the following three typical regimes:
 \begin{itemize}
\item Case 1: $\alpha > 1$, i.e.  the genetic switching process is much faster than the translation process;
\item Case 2: $\alpha = 1$, i.e.  the switching rates are comparable to the translation rates;
\item Case 3: $0<\alpha< 1$, i.e. the translation process is much faster than the genetic switching process.
\end{itemize}

In Case 2, the WKB asymptotics and the rigorous LDT results have been established for a similar model which takes into account the mRNA fluctuation\cite{LiLin}. The obtained LDT rate functional is utilized to find the most probable transition path and characterize the rate of transitions between the high and low expression states. Furthermore, the authors have shown that the Hamiltonian obtained from LDT is convex with respect to the momentum variable, which is one key point in designing robust numerical algorithms.  In Case 3, the researchers typically take the continuum limit to the translation process at first since it is even faster than the switching process\cite{Ge}. With this approach, one obtains a stochastic hybrid system which resembles similar form as those for ion channels considered in computational neuroscience. So far, the WKB asymptotics and path integral formulations are both proposed for stochastic hybrid systems. The Case 1 is also studied with WKB asymptotics applied to the averaged system with respect to the fast switching process.

From the authors' point of view, the approaches employed in [\onlinecite{Ge}] are like taking a repeated limit to the switching and translation processes according to their relative magnitudes.
\com {More concretely, when DNA switching is much faster than the protein synthesis, the equilibrium pre-averaging of the switching process is taken in [\onlinecite{Ge}] at first and one gets a pure translation process with effective translation rates; however when the protein synthesis is much faster than DNA switching, the large volume limit is taken to the translation process at first and one gets a stochastic hybrid system\cite{Ge}. Similar ideas and techniques are adopted in [\onlinecite{Newby Phys Biol}] and [\onlinecite{Newby Math Biol}]  as well, which discussed different timescale issues for the gene expression model.}  With this understanding, it will be interesting to investigate the double limit of the original process instead of taking average with respect to one faster process at first. Mathematically it is also desirable to establish the large deviations for the original system with two time scales but different magnitudes. In fact, it is the main motivation of this paper. We will utilize the Doi-Peliti second quantization path integral formalism \cite{Doi, Peliti,Wang} to study the general two-scale large deviations for the genetic switching models. As we will see, although the second quantization path integral for the spin-boson type model is formal, it is  an effective approach to derive the large deviation results for chemical jump processes. Compared with the classical path integral formalism for diffusion processes, the second quantization path integral for chemical jump processes formulates the weight of each path in an extended space which involves both coordinate and momentum variables. This makes that the large deviation result can be given through a Hamiltonian with explicit formula, which resolves the dilemma that  the Lagrangian in the rate functional does not have a closed form. This is important for further theoretical and numerical studies. Mathematically, rigorously establishing the LDT obtained from the formal approach in this paper is in progress based on our previous analysis [\onlinecite{LiLin}].

Let us briefly illustrate our general two-scale LDT results. We will show that the Lagrangian obtained from the second quantization path integral comprises of two parts, which correspond to the switching and translation processes, respectively. However, what we are interested in is the LDT only for the concentration of proteins.  The different magnitudes of the switching and translation rates essentially lead to a singularly perturbed variational problem, \com {which has different dominant terms and different scaling limits} in the cases of  $0<\alpha<1$ and $\alpha>1$. When $\alpha=1$, the Lagrangians from both parts contribute equally, and we get a result which combines the Donsker-Varadhan type LDT \cite{Varadhan} for the occupation measure of DNA states and the large volume type LDT \cite{Shwartz1995} for the small noise perturbation altogether [\onlinecite{LiLin}].  As the LDT gives the sharpest characterization of the considered two-scale chemical kinetic system, we can obtain the deterministic mean field ODEs and the chemical Langevin approximation for the system based on the local analysis of the large deviation results \cite{Dembo}. This corresponds to the law of large numbers (LLN) and the central limit theorem (CLT) for the process. We found that the three cases possess the same mean field ODEs. However, the chemical Langevin approximations for them are quite different. If $\alpha>1$, only the fluctuation from protein translation process survives. If $0<\alpha<1$, only the fluctuation from genetic switching process survives. And if $\alpha=1$, both fluctuations from protein translation and genetic switching processes contribute. Similar results are also valid for the single-molecule Michaelis-Menten kinetics \cite{Xie} with slight modifications (c.f. Section \ref{sec:Michaelis-Menten mechanism}). Our study extends the insights about the chemical kinetic systems in the classical large volume limit, and the methodology we introduced here can be applied to other multiscale problems in many fields.

The rest of the paper is organized as follows. In Section \ref{sec:two}, we introduce the chemical master equation and apply the Doi-Peliti path integral formalism to the considered model. We then rescale the system with system size $\epsilon^{-1}$ and get the abstract LDT result based on singular perturbation analysis in Section \ref{sec:three}. In Section \ref{sec:example}, we apply our abstract result to the two-state genetic switching model and present the mean field limits and chemical Langevin approximations. \com{ In Section \ref{sec:Michaelis-Menten mechanism}, we apply our result to the well-known single-molecule Michaelis-Menten kinetics and mention the implications.} Finally we make the conclusion and related discussions in Section \ref{sec:disc}.

\section{Transition Probability in a Path Integral Form}\label{sec:two}

We start from a more general model rather than Eq.~\eqref{eq:Intro1}. Assume that the DNA switching could occur among $N$ possible states ($N=2$ for the model shown in Fig. \ref{fig1}) and the chemical  master equation (CME) for the biological reaction network reads
\begin{eqnarray}
\frac{\partial}{\partial t}\Pb(n,t)&=&\Ab \Big[\Pb(n-1,t)-\Pb(n,t)\Big] \notag \\
&&+\gamma \Big[(n+1)\Pb(n+1,t)-n\Pb(n,t)\Big]\label{eq:CME} \\
&&+\Qb^{\dagger}\Pb(n,t).\notag
\end{eqnarray}
Here $\Pb(n,t)=(P_1(n,t),\ldots,P_N(n,t))^T$, $P_j(n,t)$ is the probability distribution function that the system has $n$ protein copy numbers and the switch is in state $j$ at time $t$. $\Ab$ is a diagonal matrix with diagonal entry $k_j$ as the protein synthesis rate in state $j$. $\gamma$ is the protein degradation rate and $\Qb=(q_{jk}(n))_{j,k=1}^{N}$ is the transition rate matrix among different DNA states. Thus $q_{jk}(n)\ge 0$ for any $j\neq k$ and $\sum_{k=1}^{N}q_{jk}(n)=0$. We assume that the switching process is ergodic.

Now we follow the Doi-Peliti's approach to establish the path integral formalism of the CME \eqref{eq:CME}  \cite{Wolyness, Wang, Doi,Peliti}.
Define the creation, annihilation operators $a^{\dagger}, a$ and the state function $|\psib\rangle$ as
$$a^\dag \left| n \right\rangle = \left| n+1 \right\rangle,\quad a\left| n \right\rangle = n\left| n-1 \right\rangle \text{ and }
\left| \psib  \right\rangle = \sum\nolimits_{n = 0}^\infty  \Pb(n,t) \left| n \right\rangle.$$
Then the CME \eqref{eq:CME} can be written in a second-quantized form
\begin{equation}\label{eq:2ndQF}
{\partial}_t \left| \psib \right\rangle = \Omegab \left| \psib  \right\rangle,
\end{equation}
where the operator
\begin{equation}
 \Omegab=\Ab (a^\dag-1)+\gamma(a-a^\dag a)+ \hat\Qb^{\dagger}
\end{equation}
 and $\hat\Qb$ is obtained from $\Qb$ by replacing the transition rates $q_{ij}(n)$ with operators $q_{ij}(a^\dag a)$.

\com{From Eq. \eqref{eq:2ndQF}, the transition probability $P(n_f,\tau | n_i,0)$ of finding product copy number $n_f$ at time $t=\tau$ starting from $n_i$ at $t=0$ has the form
\begin{eqnarray}\notag
P(n_f,\tau | n_i,0)&=&\left\langle {\boldsymbol n_f} \right|\exp(\Omegab \tau)\left| {\boldsymbol n_i} \right\rangle\\\label{initial P}
&=&\mathop {\lim}\limits_{\Delta t \to 0} \left\langle {\boldsymbol n_f} \right|\exp(\Omegab \Delta t)^{\tau/\Delta t}\left| {\boldsymbol n_i} \right\rangle,
\end{eqnarray}
where $\left| {\boldsymbol n_i}\right\rangle =(\left|  n_i\right\rangle,\left|  n_i \right\rangle, \cdots, \left|  n_i \right\rangle)^T$ is an $N$-dimensional column vector and $\left\langle {\boldsymbol n_f} \right|=(\left\langle { n_f} \right|, \cdots, \left\langle { n_f} \right|)$ is an $N$-dimensional row vector.
Following Zhang et al. \cite{Wang}, we utilize the coherent state representation and a resolution of identity \cite{Wolyness, Wang} as
\begin{eqnarray}\label{identity}
 \boldsymbol{I}_b \otimes\boldsymbol{I}_s=\int_0^\infty d n \int_{-\pi}^\pi\frac{d\beta}{2\pi}
 \left| z \right\rangle\left\langle {\tilde z} \right| e^{-n}\cdot
 \frac{1}{N^{N-2}}\prod\limits_{k = 1}^{N - 1}\int_0^{(N/2)\pi}\sin\frac{2\theta_k}{N}d\theta_k\frac{1}{4\pi}\int_0^{4\pi}d\phi_k| \psib^R \rangle\langle \psib^L|,
 \end{eqnarray}
where
$$| {{\psib ^R}}\rangle  = \left[ \begin{matrix}
   {\frac{2}
{N}{{\cos }^2}\frac{{{\theta _1}}}
{N}e^{i\phi_1/2}  }\\
    \vdots   \\
   {\frac{2}
{N}{{\cos }^2}\frac{{{\theta _N}}}
{N}e^{i\phi_N/2} }  \\
\end{matrix} \right], \quad \langle
\psib^L |=(e^{-i\phi_1/2},\cdots,e^{-i\phi_N/2})$$
with
$$\left| z \right\rangle=e^{a^\dag z}\left| 0 \right\rangle,\  \left\langle {\tilde z} \right|=\left\langle 0 \right|e^{a\tilde z},\ z=ne^{-i\beta},\ \tilde z=e^{i\beta},$$
 and $i$ is the imaginary unit. The variable $n$ has the interpretation that it characterizes the mean protein number in the coherent states. Define $c_j=(2/N){{\cos }^2}({\theta _j/N})$. The $c_j$ gives the
occupation probability of DNA at state $j$ from the probabilistic interpretation of quantum mechanics.
They satisfy the normalization condition $\sum_{j=1}^N c_j=1$. Correspondingly the phase variable $\phib$ can be chosen to satisfy $\sum_{j=1}^{N}\phi_j=0$ for convenience. These choices are consistent with the resolution of identity. As a consequence, we have only $N-1$ independent unknowns $(c_1,\ldots,c_{N-1})$ in
$\cb=(c_1,\ldots,c_N)$, which is equivalent to use $\thetab=(\theta_1,\ldots,\theta_N)$,  and $N-1$ unknowns $(\phi_1,\ldots,\phi_{N-1})$ in $\phib=(\phi_1,\ldots,\phi_N)$. }

Inserting \eqref{identity} into \eqref{initial P}, the transition probability density
 can be represented as a path integral form
\begin{equation}\label{transition probability}
P(n_f,\tau | n_i,0)=\text{Const.} \times \int Dn D\beta D\cb D\phib \exp \left(-\int_{0}^{\tau} L dt\right),
\end{equation}
where the Lagrangian $L$ is defined as
\begin{equation}\label{Lagrangian}
 L=i\beta\frac{dn}{dt}+\sum_{j=1}^{N-1}i\phi_j\frac{d(c_N-c_j)/2}{dt}-H(n,\cb,i\beta,i\phib),
\end{equation}
 and the Hamiltonian
\begin{equation}
 H(n,\cb,i\beta,i\phib)=H_1(n,\cb,i\beta)+H_2(n,\cb,i\phib)
\end{equation}
with $H_{1}$ for the translation process
\begin{equation}
 H_1(n,\cb,i\beta)=\sum_{j=1}^{N}k_jc_j[\exp(i\beta)-1]+\gamma n[\exp(-i\beta)-1]
\end{equation}
and $H_{2}$ for the switching process
\begin{equation}
 H_2(n,\cb,i\phib) = \sum_{m,j}^{N}c_m q_{mj}(n)(e^{(i\phi_m-i\phi_j)/2}-1).
\end{equation}
\com{In Eq. \eqref{transition probability}, the outer 4-fold integral is taken in the path space with respect to $n(t),\beta(t),\cb(t)$ and $\phib(t)$, which are full trajectories in $[0,\tau]$. The terms involving $| n_i\rangle$ and $\langle n_f|$ have been absorbed to the constant before the integral. The path integral formulation \eqref{transition probability} makes the weight of each trajectory explicit. The form of Lagrangian \eqref{Lagrangian} suggests the  interpretation that the pairs $i\beta$ and $n$, $i\phi$ and $c$ are conjugate variables.}

To study the associated LDT, we must have a small parameter $\epsilon$ and a deterministic limit as $\epsilon\rightarrow 0$. This could be chosen as the inverse of typical system size $\epsilon=\gamma/k_{1}=V^{-1}$.
As stated in the introduction section,  we assume
\begin{eqnarray}\label{eq:scaling}
 k_j \sim \frac{1}{\epsilon},\quad q_{ij} \sim \frac{1}{\epsilon^\alpha}, \quad \alpha>0
\end{eqnarray}
and define
\begin{equation}\label{eq:VarTilde}
x=n\epsilon,\quad\tilde k_j=k_j\epsilon,\quad \tilde q_{jk}(x)= q_{jk}(n)\epsilon^\alpha.
\end{equation}
With these definitions, Eq. \eqref{transition probability} can be rewritten as
\begin{equation}
P(n_f,\tau | n_i,0)=\text{Const.} \times \int Dx D\beta D\cb D\phib \exp \left(-\frac{1}{\epsilon}\int _{0}^{\tau}\tilde L_1 dt -\frac{1}{\epsilon^\alpha}\int_{0}^{\tau} \tilde L_2 dt \right),
\end{equation}
where the rescaled Lagrangian
\begin{eqnarray}
 &&\tilde L_1= i\beta\frac{dx}{dt}-\tilde H_1(x,\cb,i\beta),\label{eq:L1tilde}\\
 &&\tilde L_2= \epsilon^\alpha\sum_{j=1}^{N-1}i\phi_j\frac{d(c_N-c_j)/2}{dt}-\tilde H_2(x,\cb,i\phib), \label{eq:L2tilde}
\end{eqnarray}
and rescaled Hamiltonian
\begin{eqnarray}\label {H 1}
 &&\tilde H_1(x,\cb,i\beta)= \sum_{j=1}^{N}\tilde k_jc_j[\exp(i\beta)-1]+\gamma x[\exp(-i\beta)-1],\\ \label{H 2}
 &&\tilde H_2(x,\cb,i\phib)= \sum_{m,j}^{N}c_m \tilde q_{mj}(x)(e^{(i\phi_m-i\phi_j)/2}-1) .
\end{eqnarray}
Using the method of steepest descent asymptotics, the integration over $\beta$ and $\phib$ can be approximated by simply using the value of the integrand at the saddle point \cite{Orszag}. Thus, we get
\begin{equation}\label{eq:TransitionProbability2}
 P(n_f,\tau | n_i,0) \propto \int Dx  D\cb \exp \left(-\frac{1}{\epsilon}\int_{0}^{\tau}  L_1(x,\dot x,\cb) dt-\frac{1}{\epsilon^\alpha}\int_{0}^{\tau} L_2(x,\cb)dt\right),
\end{equation}
where
\begin{eqnarray}
L_1(x,\dot x,\cb)&=&\mathop{\sup}_p\{p \dot x-\tilde H_1(x,\cb,p)\},\label{eq:L1}\\
L_2(x, \cb)&=&\mathop{\sup}_{\varphib} \{-\tilde H_2(x,\cb,\varphib)\}.\label{eq:L2}
\end{eqnarray}
Note the term  $\varphib\cdot\dot{\cb}$ does not appear in Eq. \eqref{eq:L2} because of the factor $\epsilon^{\alpha}$ in the first term of Eq. \eqref{eq:L2tilde}. \com {
Formally, the functional appearing in the exponential in \eqref{eq:TransitionProbability2} is a competition between the rate functional $\int L_1dt $ which corresponding to the translation process and the rate functional $\int  L_2dt$ which corresponding to the switching process. It is interesting to observe that the Lagrangian $L_1$ corresponds to the large volume type LDT rate function for the small noise perturbation  \cite{Shwartz1995} and $L_2$ corresponds to the Donsker-Varadhan type LDT rate function for the occupation measure of DNA states \cite{Shwartz1995,Varadhan}.  The second quantization path integral perfectly reveals the intrinsic structure of the considered two-scale chemical kinetic process.}

\section{Formulation of the LDT in a general setting}\label{sec:three}

The transition probability \eqref{eq:TransitionProbability2} contains the LDT information about the variables $x$ and $\cb$. However in most cases, one is only interested in slow variables, i.e. the
concentration of protein in our case, which is also the observable in experiments.
 In this sense, we must integrate over $\cb$-space. It turns out the final result depends on the value of $\alpha$ and we will have three typical regimes. In what follows, we will discuss different outcomes in different regimes separately.

 \vspace*{2mm}
 {\bf (i). Case 1: $\alpha > 1$.}  The switching process is much faster than the translation process.

 \vspace*{2mm}
In this case, we can rewrite Eq. \eqref{eq:TransitionProbability2} as
\begin{equation}
 P(n_f,\tau | n_i,0) \propto \int Dx  D\cb \exp \left\{-\frac{1}{\epsilon}\int_{0}^{\tau} dt \left(L_1(x,\dot x,\cb)+\frac{1}{\epsilon^{\alpha-1}} L_2(x,\cb)\right)\right\}.
\end{equation}
To integrate over $\cb$-space, we take the Laplace asymptotics for each $t$. The Lagrangian for $x$ has the form
\begin{eqnarray}\label{eq:Lx1}
L_x(x,\dot x)=\mathop{\inf}_{\cb}\left\{L_1(x,\dot x,\cb)+\frac{1}{\epsilon^{\alpha-1}}L_2(x,\cb)\right\}\quad\text{as}\quad \epsilon\rightarrow 0+.
\end{eqnarray}
Since $L_2 \ge 0$ and $\epsilon^{1-\alpha}\gg 1$, the term $\epsilon^{1-\alpha}L_2(x,\cb)$ dominates. From the assumption that the switching process is ergodic, for a given $x$, $L_2(x,\cb)$ achieves its minimum $L_2(x,\cb)=0$ at a single point $\cb=\cb_0(x)$, i.e. the steady state distribution given the concentration $x$ \cite{Shwartz1995,Kemeny}. Thus we get
\begin{eqnarray}\label{eq:LxCon}
L_x(x,\dot x)&=&L_1(x,\dot x,\cb_0(x))
\end{eqnarray}
and
\begin{equation}
P(n_f,\tau | n_i,0) \propto \int Dx \exp\left\{{-\frac{1}{\epsilon}\int_{0}^{\tau} dt L_x(x,\dot x)}\right\}.
\end{equation}
Although we still leave the factor $\epsilon^{1-\alpha}$ in the Laplace aysmptotics \eqref{eq:Lx1} in our manipulation and then take the singular perturbation analysis, it is not difficult to establish the final result in a rigorous way. This result tells us that when $\alpha>1$ the LDT for the slow variable
 $x$ is only determined by the effective synthesis rate $\sum_{j=1}^{N}\tilde k_j({\cb}_{0}(x))_j$ and degradation rate $\gamma$.

\vspace*{2mm}
{\bf  (ii). Case 3: $0<\alpha< 1$.} The translation process is much faster than the switching process.

\vspace*{2mm}
In this case, we rewrite Eq. \eqref{eq:TransitionProbability2} as
\begin{equation}
 P(n_f,\tau | n_i,0)
 \propto \int Dx  D\cb \exp \left\{-\frac{1}{\epsilon^\alpha}\int dt \left(\frac{1}{\epsilon^{1-\alpha}}  L_1(x,\dot x,\cb)+ L_2(x,\cb)\right)\right\}.
\end{equation}
Taking Laplace asymptotics with respect to $\cb$-integral, we get
\begin{eqnarray}\label{Lx x}
L_x(x,\dot x)=\mathop{\inf}_{\cb}\left(\frac{1}{\epsilon^{1-\alpha}} L_1(x,\dot x,\cb)+ L_2(x,\cb)\right).
\end{eqnarray}
Since $L_1 \ge 0$ and  $\epsilon^{\alpha-1}\gg 1$, the term $\epsilon^{\alpha-1} L_1(x,\dot x,\cb)$ dominates. We can perform similar approach to derive $L_{x}$ as in the previous case. In general, we assume that $L_1(x,\dot x,\cb)$ achieves its minimum $L_1(x,\dot x,\cb)=0$ at $\cb=\cb_x$ for a given $x$. In our case, $\cb_x$ satisfies the mean field ODE by large volume limit:
 \begin{equation}\label{xs}
 \dot x =\sum_{j=1}^{N} k_j (\cb_{x})_{j} - \gamma x.
\end{equation}
By \eqref{Lx x} and \eqref{xs}, we have
\begin{eqnarray}\label{eq:Lx2}
L_x(x,\dot x)=\mathop{\inf}_{\{\cb_x:\dot x =\sum_{j=1}^{N} k_j( \cb_{x})_{j} - \gamma x\}}L_2(x,\cb_x)
\end{eqnarray}
and
\begin{equation}\label{eq:PdfCase3}
P(n_f,\tau | n_i,0) \propto \int Dx \exp\left\{{-\frac{1}{\epsilon^\alpha}\int_{0}^{\tau} dt L_x(x,\dot x)}\right\}.
\end{equation}
We want to remark here that from \eqref{eq:PdfCase3} we will expect to get the LDT of the type
\begin{equation}\label{eq:LDTCase3}
\lim_{\epsilon\rightarrow 0+}\epsilon^{\alpha}\ln \Prob(X_{\cdot}\in \mathcal{B}) = -\inf_{x\in \mathcal{B}} \int_{0}^{\tau} L_{x}(x,\dot{x})dt
\end{equation}
where $\mathcal{B}$ is a Borel set in $D[0,\tau]$ space (functions on $[0,\tau]$ are right continuous with left limits) and $X_{\cdot}$ is the sample path of the original jump process. The scaling $\epsilon^{\alpha}$  in \eqref{eq:LDTCase3} is essential to reveal the nontrivial behavior of $x$. Other choices of the exponent do not give the correct limit which we are interested in for $x$.

In the considered case $0<\alpha<1$, the protein synthesis are much faster than the genetic switching. With this condition, if we neglect the copy-number
fluctuation of the protein, we get a reduced stochastic hybrid system:
\begin{equation}\label{eq:HybridSystem}
\dot x= \sum_{j=1}^N k_j I_{\{\xi(t)=j\}} - \gamma x,
\end{equation}
where $I_{\{\xi(t)=j\}}$ is an indicator function and $\xi(t)$ represents the DNA occupation state.
In [\onlinecite{Bressloff}] and [\onlinecite{FAGGIONATO}], the authors established the LDTs for variable $x$ as $\epsilon \to 0+$ for the system \eqref{eq:HybridSystem}, which is the same as what we derived in Eq. \eqref{eq:Lx2}. But we should emphasize that this coincidence is not obvious {\it a priori}, our result supports the validity of the procedure  by taking the repeated limit for  two-scale processes in some sense.

\vspace*{2mm}
{\bf  (iii). Case 2: $\alpha = 1$.}  The switching rates are comparable to the translation rates.

\vspace*{2mm}
When $\alpha=1$, we have
\begin{equation}
P(n_f,\tau | n_i,0)
 =\int Dx  D\cb \exp \left\{-\frac{1}{\epsilon}\left(\int dt L_1(x,\dot x,\cb)+\int dt L_2(x,\cb)\right)\right\}.
\end{equation}
In this case, we have the LDT Lagrangian for variable $x$:
\begin{eqnarray} \label{eq:Lx3}
L_x(x,\dot x)=\mathop{\inf}_{\cb}\left\{L_1(x,\dot x,\cb)+L_2(x,\cb)\right\}.
\end{eqnarray}
Since in most cases there is no closed form for $L_{1}$, thus we do not expect to get the closed form of $L_{x}$ accordingly. This hinders the applicability of the obtained theory. It is more convenient to study the conjugate Hamiltonian of $L_{x}$:
\begin{eqnarray}\notag
H_x(x,p)&=&\mathop{\sup}_{\beta}\left\{p\beta- L_x(x,\beta)\right\}\\\notag
&=&\mathop{\sup}_{\beta}\left\{p\beta- \mathop{\inf}_{\cb}\left\{L_1(x,\beta,\cb)+L_2(x,\cb)\right\}\right\}\\\notag
&=&\mathop{\sup}_{\beta}\mathop{\sup}_{\cb}\left\{p\beta- L_1(x,\beta,\cb)-L_2(x,\cb)\right\}\\\notag
&=&\mathop{\sup}_{\cb}\mathop{\sup}_{\beta}\left\{p\beta- L_1(x,\beta,\cb)-L_2(x,\cb)\right\}\\\label{eq:Hx3}
&=&\mathop{\sup}_{\cb}\left\{\tilde H_1(x,p,\cb)-L_2(x,\cb)\right\}.
\end{eqnarray}
As we will show, the dual Hamiltonian may have explicit expression and it is convex with respect to the momentum variable $p$. This property makes it competitive for the numerical algorithms for solving static Hamilton-Jacobi equation through the geometric minimum action method (gMAM)\cite{gMAM, Lv14}.

\section{Application to the two-state model}\label{sec:example}

Using the two-state model \eqref{eq:Intro1} as an example, we will give the detailed LDT results for different $\alpha$, and show the mean field ODE and the chemical Langevin approximation for variable $x$. Moreover, we will solve the static Hamilton-Jacobi equation for the quasi-potential $\Phi(x)$ in different situations. At first, we take the same rescaling \eqref{eq:VarTilde} for the variables and parameters.  We again consider three different cases: (i) $\alpha>1$, (ii) $0<\alpha<1$ and (iii) $\alpha=1$.

\vspace*{2mm}
 {\bf (i). Case 1: $\alpha > 1$.}

\vspace*{2mm}
In this case, the ergodic limit of DNA occupation probability is
$$\cb_0(x)=\left(\frac{\tilde{f}(x)}{\tilde{f}(x)+\tilde{g}(x)},~\frac{\tilde{g}(x)}{\tilde{f}(x)+\tilde{g}(x)}\right)$$
for given $x$. By Eq. \eqref{eq:Lx1}, we have
\begin{equation}
L_x(x,\dot x)=\mathop{\sup}_p \left\{p \dot x-\left[\frac{\tilde k_1 \tilde f(x)+\tilde k_0 \tilde g(x)}{\tilde f(x)+\tilde g(x)}(\exp(p)-1)+\gamma x(\exp(-p)-1)\right]\right\},
\end{equation}
and the dual Hamiltonian
\begin{equation}\label{Hx ex 1}
H_x(x,p)=\frac{\tilde k_1\tilde f(x)+\tilde k_0 \tilde g(x)}{\tilde f(x)+\tilde g(x)}[\exp(p)-1]+\gamma x[\exp(-p)-1].
\end{equation}

From the result
\begin{equation}
\frac{\partial H_x}{\partial p} \Big|_{p = 0} = \frac{\tilde k_1\tilde f(x)+\tilde k_0g(x)}{\tilde f(x)+ \tilde g(x)}-\gamma x,
\end{equation}
we get the mean field ODE
\begin{equation}
\frac{dx}{dt}=\frac{\partial H_x}{\partial p} \Big|_{p = 0} = \frac{\tilde k_1\tilde f(x)+\tilde k_0\tilde g(x)}{\tilde f(x)+\tilde g(x)}-\gamma x.
\end{equation}
Furthermore, the fact
\begin{equation}
\frac{\partial ^2 H_x}{\partial p^2} \Big|_{p = 0} =  \frac{\tilde k_1\tilde f(x)+\tilde k_0\tilde g(x)}{\tilde f(x)+\tilde g(x)}+\gamma x
\end{equation}
shows the following chemical Langevin approximation holds
\begin{equation}\label{eq:LangevinApproximation1}
dx=\left(\frac{\tilde k_1f(x)+\tilde k_0\tilde g(x)}{\tilde f(x)+\tilde g(x)}-\gamma x\right)dt+\sqrt{\epsilon}\left(\sqrt{\frac{\tilde k_1 \tilde f(x) +\tilde k_0 \tilde g(x)}{\tilde f(x)+\tilde g(x)}}dw_1-\sqrt{\gamma x}dw_2\right),
\end{equation}
where $w_1$ and $w_2$ are independent standard Brownian motions.

From classical variational analysis \cite{Rockafellar}, it can be shown that the quasi-potential
defined through
$$
\Phi(x;x_{0}) = \inf_{\tau\ge 0}\inf_{x(0)=x_{0},x(\tau)=x}\int_{0}^{\tau} L_{x}(x,\dot{x})dt
$$
in our case satisfies a static Hamilton-Jacobi equation
$H(x,\partial_x \Phi ) = 0$, where $x_{0}$ is a stable fixed point.  Based on \eqref{Hx ex 1}, we have  by some algebra
\begin{equation}\label{quasipotential 1}
\partial_x \Phi= - \log \frac{(\tilde k_1 \tilde f(x)+\tilde k_0\tilde g(x))/(\tilde f(x)+\tilde g(x))}{\gamma x}.
\end{equation}
This result is consistent with the quasi-potential derived in [\onlinecite{Ge}], where the authors neglect the fluctuation of genetic switching and get the result by WKB ansatz. But of course, there is no hope to get the explicit formula of $\Phi$ when the dimension of $x$ is bigger than 1.

\vspace*{2mm}
 {\bf (ii). Case 3: $0<\alpha<1$.}

\vspace*{2mm}
In this case, we have the Lagrangian
\begin{eqnarray}\notag
L_x(x,\dot x)&=&L_2(x,\cb_x)\\\notag
&=&\mathop{\sup}_\varphi \left\{-\left[c_1 \tilde g(x)(e^{\phi_1-\phi_2}-1) + c_2 \tilde f(x)(e^{\phi_2-\phi_1}-1)\right]\right\}\\
&=& \left(\sqrt{\tilde f(x) c_2} - \sqrt{\tilde g(x)c_1}\right)^2,
\end{eqnarray}
where $\cb_{x}=(c_{1},c_{2})$ and $c_1=(\dot x - \tilde k_0 + \gamma x)/(\tilde k_1 - \tilde k_0)$, $c_2=1-c_1$ by the condition $L_{1}(x,\dot{x},\cb_{x})=0$.
With the Legendre-Fenchel transform defined by $H_x(x,p)
=\mathop{\sup}_\beta\left(p\beta-L_x(x,\beta)\right)$,  we get the dual
 Hamiltonian:
\begin{equation}\label{eq:HxEx2}
H_x(x,p)=p\beta_0-L_x(x,\beta_0),
\end{equation}
where $\beta_0=\tilde{k}_1 s_1+\tilde{k}_2 (1-s_1)- \gamma x$ and
$$s_1=\frac{1}{2}+\frac{s_2}{2\sqrt{s_2^2+4}}, \quad
s_2=\frac{p(\tilde k_1- \tilde k_0)+\tilde f(x)-\tilde g(x)}{\sqrt{\tilde f(x) \tilde g(x)}}.$$

Again, we can obtain the deterministic mean field ODE as
\begin{equation}
\frac{dx}{dt}=\frac{\partial H_x}{\partial p} \Big|_{p = 0}=\frac{\tilde k_1\tilde f(x)+\tilde k_0\tilde g(x)}{\tilde f(x)+\tilde g(x)}-\gamma x.
\end{equation}
Similarly, we get
\begin{equation}
\frac{\partial ^2 H_x}{\partial p^2} \Big|_{p = 0} =  \frac{2\tilde f(x)\tilde g(x)}{(\tilde f(x)+\tilde g(x))^3}(\tilde k_1-\tilde k_0)^2
\end{equation}
and thus the chemical Langevin approximation
\begin{equation}\label{Langevin approximation 2}
dx=\left(\frac{\tilde k_1\tilde f(x)+\tilde k_0\tilde g(x)}{\tilde f(x)+\tilde g(x)}-\gamma x\right)dt+\sqrt{\epsilon^\alpha}\sqrt{\frac{2\tilde f(x)\tilde g(x)}{(\tilde f(x)+\tilde g(x))^3}(\tilde k_1-\tilde k_0)^2}dw.
\end{equation}
We note here that the fluctuation term has the strength $\sqrt{\epsilon^{\alpha}}$ since it originates from the fast genetic switching process, and the term $\sqrt{\gamma x}dw_{2}$ disappears because it is in order $\sqrt{\epsilon}$. These are in sharp contrast with the result in \eqref{eq:LangevinApproximation1} and Case 2 below which has the $O(\sqrt{\epsilon})$ fluctuation.

By Eq. \eqref{eq:HxEx2}, solving the Hamilton-Jacobi equation $H(x,\partial_x \Phi ) = 0$, we have
\begin{equation}\label{quasipotential 2}
\partial_x \Phi= \frac{\tilde f}{\tilde k_0 -\gamma x}+ \frac{\tilde g}{\tilde k_1 -\gamma x}.
\end{equation}
This is consistent with the result in [\onlinecite{Ge}] although we have totally different form of Hamiltonian $H$.

\vspace*{2mm}
 {\bf (iii). Case 2: $\alpha=1$.}

\vspace*{2mm}
In this case, the genetic switching rate is comparable to the protein synthesis rate. The rigorous LDT result has been obtained in  [\onlinecite{LiLin}]. Now we formally establish the LDT again through the  second-quantization path integral approach.

By \eqref{eq:Lx3} and \eqref{eq:Hx3}, we have the dual Hamilton:
\begin{eqnarray}\notag
H_x(x,p)&=&\mathop{\sup}_{\cb}\left\{\tilde H_1(x,\cb,p)-L_2(x,\cb)\right\}\\
        &=&\mathop{\sup}_{\cb}\bigg\{\left(\tilde k_1 c_1+\tilde k_0 c_2\right)[\exp(p)-1]+\gamma x[\exp(-p)-1]-\left(\sqrt{(\tilde f(x) c_2}-\sqrt{\tilde g(x) c_1}\right)^2\bigg\}.
\end{eqnarray}
Since $c_1+c_2=1$ and $c_{j}\ge 0$, we can obtain the explicit expression of $H_x(x,p)$:
\begin{eqnarray}\label{H alpha=1}
H_x(x,p)&=&\left(\tilde k_1 s+\tilde k_0(1-s)\right)[\exp(p)-1]+\gamma x[\exp(-p)-1]-\left(\sqrt{(\tilde f(1-s)}-\sqrt{\tilde gs}\right)^2.
\end{eqnarray}
where
$$s=\frac{1}{2}+\frac{s_1}{2\sqrt{s_1^2+4}},\quad s_1=\frac{(\tilde k_1-\tilde k_0)(e^p-1)+\tilde f(x)-\tilde g(x)}{\sqrt{\tilde f(x) \tilde g(x)}}.$$

As before, we get the mean field ODE
\begin{equation}
\frac{dx}{dt}=\frac{\partial H_x}{\partial p} \Big|_{p = 0}=\frac{\tilde k_1\tilde f(x)+\tilde k_0\tilde g(x)}{\tilde f(x)+\tilde g(x)}-\gamma x.
\end{equation}
The second order expansion to $p$
\begin{equation}
\frac{\partial ^2 H_x}{\partial p^2} \Big|_{p = 0} =  \frac{2\tilde f(x)\tilde g(x)}{(\tilde f(x)+\tilde g(x))^3}(\tilde k_1-\tilde k_0)^2+\frac{\tilde k_1\tilde f(x)+\tilde k_0\tilde g(x)}{\tilde f(x)+\tilde g(x)}+\gamma x
\end{equation}
yields the following chemical Langevin approximation
\begin{eqnarray}\label{Langevin approximation 3}
dx&=&\left(\frac{\tilde k_1\tilde f+\tilde k_0\tilde g}{\tilde f+\tilde g}-\gamma x\right)dt+\sqrt{\epsilon}\left(\sqrt{\frac{\tilde k_1\tilde f+ \tilde k_0\tilde g}{\tilde f+\tilde g}+\frac{2\tilde f\tilde g}{(\tilde f+\tilde g)^3}(\tilde k_1-\tilde k_0)^2}dw_1-\sqrt{\gamma x}dw_2\right).
\end{eqnarray}
where $\tilde{f},\tilde{g}$ are abbreviations of $\tilde{f}(x)$ and $\tilde{g}(x)$, and $w_1,w_2$
are independent standard Brownian motions.

\com{It is worth discussing the relationship between the Hamiltonian \eqref{H alpha=1} and that obtained by WKB asymptotics. To get a Hamiltonian via WKB asymptotics, we follow the  procedures in [\onlinecite{Newby Dyn Syst}] and sketch its outline. We assume that the
stationary solution of \eqref{eq:CME} has the form
\begin{equation}\label{stationary distribution}
P_i^s(x)\sim r_{i}(x)k(x)\exp\left[-\frac{1}{\epsilon}\Phi(x)\right], \quad i=1,2.
\end{equation}
Substituting \eqref{stationary distribution} into \eqref{eq:CME} and collecting leading order terms, we get $M(x,p)\cdot(r_{1}(x), r_{2}(x))^T=0$, where
$$M(x,p)=\left(
\begin{array}{cc}
\tilde k_1(e^p-1)+\gamma x(e^{-p}-1)-\tilde g & \tilde f  \\
 \tilde g & \tilde k_0(e^p-1)+\gamma x(e^{-p}-1)-\tilde f
\end{array}
\right).$$
Now there is a subtlety to get the correct Hamilton in LDT. If one takes the choice that the $H(x,p)$ is defined as the largest eigenvalue of $M(x,p)$, it can be shown that it is equivalent to \eqref{H alpha=1}. However if one takes another choice that it is defined as the determinant of $M(x,p)$ \cite{Newby Dyn Syst, Assaf},  then $H$ is not convex in momentum variable $p$ and the equivalence is lost.  This resembles the issue of the choice of Hamiltonians for parametrized curve problem in classical mechanics (see p. 40 in [\onlinecite{Buhler}]).}

\com{Let us make some comments on the obtained mean-field limit and Langevin approximations. Recall that there are two parts of noise in the original dynamics: one is from the translation process and the other from DNA switching process.  Our result tells that different diffusion approximations arise according to the magnitudes of residual noise in different reaction channels. If $\alpha>1$, the dominant part of noise is from the translation, so only the fluctuation from translation process survives. If $0<\alpha<1$, the dominant part is from switching process, so only the fluctuation from DNA switching process survives. And if $\alpha=1$, both fluctuations from protein translation and genetic switching contribute. Similar situation occurs in the LDT analysis, where the singular perturbation is performed for variational minimizations. The obtained results show the validity of the procedure  by taking the limit for faster process at first and then performing the corresponding analysis for slower process. Although we only consider the two-state models, the essential structure and results hold for general cases. It is a natural extension of the classical large volume limit for chemical reaction processes. We summarize our discussions for the three regimes in  Table \ref{comparison table}.}
\renewcommand{\arraystretch}{2}
\begin{table}[htbp]
\centering
\caption{\com{Comparison of LDTs, mean-field limits and Langevin approximations for three regimes}}\label{comparison table}
\begin{tabularx}{0.93\textwidth}{c |c|c|c}
\Xhline{1.2pt}\hline
    &\com{ LDT Hamiltonian }&\com{ Deterministic Drift}  & \com{Noise in Langevin Approximation}  \\ \Xhline{1.2pt}
 {\com{\bf Case 1:}} \com{($\alpha>1$)} & \com{$H_x(x,p)$ in \eqref{Hx ex 1}} & \multirow{3}[15]{*}{\com{$\displaystyle \frac{\tilde k_1\tilde f(x)+\tilde k_0\tilde g(x)}{\tilde f(x)+\tilde g(x)}-\gamma x$}} & \Gape[1.2mm]{\com{$\displaystyle\sqrt{\epsilon}\left(\sqrt{\frac{\tilde k_1 \tilde f +\tilde k_0 \tilde g}{\tilde f+\tilde g}}dw_1-\sqrt{\gamma x}dw_2\right)$}} \\ \cline{1-2}\cline{4-4}
    {\com{\bf Case 2:}} \com{ ($\alpha=1$)}  & \com{$H_x(x,p)$ in \eqref{H alpha=1}}  &    &  \Gape[1.2mm]{\com{$\displaystyle\sqrt{\epsilon}\left(\sqrt{\frac{\tilde k_1\tilde f+ \tilde k_0\tilde g}{\tilde f+\tilde g}+\frac{2\tilde f\tilde g}{(\tilde f+\tilde g)^3}(\tilde k_1-\tilde k_0)^2}dw_1-\sqrt{\gamma x}dw_2\right)$}} \\ \cline{1-2}\cline{4-4}
    {\com{\bf Case 3:}} \com{($\alpha<1$)}  & \com{$H_x(x,p)$ in \eqref{eq:HxEx2}} &  & \Gape[1.2mm]{\com{$\displaystyle\sqrt{\epsilon^\alpha}\sqrt{\frac{2\tilde f\tilde g}{(\tilde f+\tilde g)^3}(\tilde k_1-\tilde k_0)^2}dw_1$ }}  \\ \Xhline{1.2pt}
    \end{tabularx}%
\end{table}%

\section{Application to the Single-Molecule Michaelis-Menten kinetics}\label{sec:Michaelis-Menten mechanism}

\com{Our approach and observation have interesting implications on the  single-molecule Michaelis-Menten system \cite{Xie}, in which a substrate S binds reversibly with an enzyme E to form an enzyme-substrate
complex ES that decomposes to form a product P. The reaction schemes can be schematically shown as
\begin{eqnarray}\label{MMM}
\begin{array}{ccccc}
{\rm E+S} & \autorightleftharpoons{$k_1$}{$k_{-1}$} & ${{\rm ES}}$ & \autorightarrow{$k_2$}{} & ${{\rm E+P}}$,
\end{array}\quad \quad
\begin{array}{ccc}
{\rm P}& \autorightarrow{${\tiny\gamma }$}{} & \emptyset.
\end{array}
\end{eqnarray}
In case of single-molecule enzyme set-up, the reaction system \eqref{MMM} falls in the framework considered in this paper. As in [\onlinecite{Xie}], we assume that the substrate is abundant enough and  there is essentially no depletion of substrate by a single enzyme molecule. That is, we assume the concentration of substrate is a constant, which will be denoted as $[S]$. It is well-known that the rate of product formulation $v$ has the following form in the quasi-steady state approximation \cite{Keener}
\begin{equation}\label{MMODE}
 v=\frac{k_2[S]}{[S]+k_M},
\end{equation}
where $k_M=(k_{-1}+k_2)/k_1$. In [\onlinecite{Xie}],  the statistics of enzymatic turn-over time and dynamical disorder are considered. Here we are interested in deriving the Langevin approximations of the Michaelis-Menten system in different regimes.

In [\onlinecite{Xie}], $k_{-1}$ ranges from $0s^{-1}$ to $2000s^{-1}$, $k_1$ is usually taken as $10^7M^{-1}s^{-1}$, $k_2=250s^{-1}$, and $[S]$ ranges from the order $0.001mM$ to $0.1mM$, where $1M=1\text{mol/L}$. Some specific choices of these parameters include
\begin{itemize}
\item Case 1: $ k_{-1}=2000s^{-1}, k_1[S]=10^7M^{-1}s^{-1}\times 0.30mM=3000s^{-1}, k_2=250s^{-1};$
\item Case 2: $ k_{-1}=200s^{-1}, k_1[S]=10^7M^{-1}s^{-1}\times 0.02mM=200s^{-1}, k_2=250s^{-1};$
\item Case 3: $ k_{-1}=50s^{-1}, k_1[S]=10^7M^{-1}s^{-1}\times 0.005mM=50s^{-1}, k_2=250s^{-1}.$
\end{itemize}
The degradation rate constant $\gamma$ is not essential and we assume it is $O(1s^{-1})$. The above choices underlie the rationale to  study different regimes in previous sections since we can make the assumption
\begin{eqnarray}\label{eq:ParaScaling}
 k_2 \sim \frac{1}{\epsilon},\quad  k_1[S], k_{-1} \sim \frac{1}{\epsilon^\alpha}, \quad \alpha>0
\end{eqnarray}
if we define $\epsilon=1/250$.

The chemical master equation of the system \eqref{MMM} can be written as
\begin{eqnarray}
{\partial_t}\Pb(n,t)&=&\left(
\begin{array}{cc}
0 & k_2  \\
 0 & 0
\end{array}
\right) (\Eb_{n}^{-1}-1)\Pb(n,t)+\gamma (\Eb_{n}^{1}-1)[n\Pb(n,t)]+
\left( {\begin{array}{cc}
   { -k_1[S]} & {k_{-1}+k_2}  \\
   {k_1[S]} & { -(k_{-1}+k_2)}  \\
\end{array}} \right)
\Pb(n,t).
\end{eqnarray}
Denote $c_1$ the occupation probability of the free enzyme molecule state E and $c_2$ the probability of the complex state ES. With similar approach in deriving \eqref{transition probability}, the transition probability can be obtained  with Lagrangian $L$
\begin{equation}
 L=i\beta\frac{dn}{dt}+i\phi_1\frac{d(- c_1)}{dt}-H(n,\cb,i\beta,i\phib),
\end{equation}
where the Hamiltonian
\begin{equation}
 H(n,\cb,i\beta,i\phib)=k_2c_2[\exp(i\beta)-1]+\gamma n[\exp(-i\beta)-1]+c_2(k_{-1}+k_2e^{i\beta})(e^{i\phi_2}-1)+c_1k_1[S](e^{-i\phi_2}-1).
\end{equation}

According to \eqref{eq:ParaScaling}, we make the rescaling
\begin{eqnarray}
 x=n\epsilon\ (\text{or} \ x_\alpha = n\epsilon^\alpha),\quad  \tilde k_2 = k_2{\epsilon},\quad  \tilde k_1= k_1[S]{\epsilon^\alpha},  \quad  \tilde k_{-1}= k_{-1}{\epsilon^\alpha}.
\end{eqnarray}
Then the Cases 1, 2 and  3 correspond to $\alpha>1$, $\alpha=1$ and $0<\alpha<1$, respectively.
Next let us study the three cases separately. The order of discussion will be from easy to difficult, which may be slightly different from previous sections.

\vspace*{2mm}
 {\bf (i). Case 2:} $\alpha=1$.

 \vspace*{2mm}
With the steepest descent asymptotics as in \eqref{eq:TransitionProbability2}, we have
 \begin{equation}\label{eq:TransitionPDF}
 P(n_f,\tau | n_i,0) \propto \int Dx  D\cb \exp \left(-\frac{1}{\epsilon}\int_{0}^{\tau}  L(x,\dot x,\cb) dt\right).
\end{equation}
The Lagrangian $L$ has the form
\begin{eqnarray}\notag
L(x,\dot x,\cb)&=&\mathop{\sup}_p\mathop{\sup}_\varphi\{p \dot x-\tilde H(x,\cb,p,\varphi)\}\\
&=& \mathop{\sup}_p\{p \dot x-H(x,\cb,p)\},\label{eq:MML}
\end{eqnarray}
where $H(x,\cb,p)=\inf_\varphi\{\tilde H(x,\cb,p,\varphi)\}$ and
\begin{eqnarray}
\tilde H(x,\cb,p,\varphi)=\tilde k_2c_2(e^p-1)+\gamma x(e^{-p}-1)+c_2(\tilde k_{-1}+\tilde k_2e^p)(e^\varphi-1)+c_1\tilde k_1(e^{-\varphi}-1).
\end{eqnarray}
In this case, we have the LDT Lagragian for variable $x$ by applying Laplace asymptotics
\begin{eqnarray}
L_x(x,\dot x)=\mathop{\inf}_{\cb}\left\{L(x,\dot x,\cb)\right\}.
\end{eqnarray}
Following the approaches in deriving \eqref{eq:Hx3}, we get the conjugate Hamiltonian of $L_x$:
 \begin{eqnarray}\notag
 H_x(x,p)&=&\mathop{\sup}_{\beta}\left\{p\beta- L_x(x,\beta)\right\}=\mathop{\sup}_{\beta}\left\{p\beta- \mathop{\inf}_{\cb}\left\{L(x,\beta,\cb)\right\}\right\}\\\notag
&=&\mathop{\sup}_{\beta}\mathop{\sup}_{\cb}\left\{p\beta- L(x,\beta,\cb)\right\}=\mathop{\sup}_{\cb}\mathop{\sup}_{\beta}\left\{p\beta- L(x,\beta,\cb)\right\}\\
&=&\mathop{\sup}_{\cb}\left\{H(x,\cb,p)\right\}=\sup_{\cb}\inf_\varphi\{\tilde H(x,\cb,p,\varphi)\}\label{eq:MMH}\\
 &=&\tilde k_2 s(e^p-1)+\gamma x(e^{-p}-1)-\left(\sqrt{\tilde k_1(1-s)}-\sqrt{ (\tilde k_{-1}+\tilde k_2e^p) s}\right)^2,\label{eq:MMCase2H}
\end{eqnarray}
where
$$s=\frac{1}{2}+\frac{s_1}{2\sqrt{s_1^2+4}},\quad s_1=\frac{\tilde k_1-(\tilde k_{-1} + \tilde k_2)}{\sqrt{\tilde k_1(\tilde k_{-1}+ \tilde k_2e^p) }}.$$

We have the mean field ODE by local analysis
 \begin{equation}\label{eq:MMCase2}
\frac{dx}{dt}=\frac{\partial H_x}{\partial p} \Big|_{p = 0}=\frac{\tilde k_1\tilde k_2}{\tilde k_1+\tilde k_{-1} + \tilde k_2}-\gamma x.
\end{equation}
This is consistent with the Michaelis-Menten law shown in \eqref{MMODE}. To see this, we first note that the reaction rate $v$ should be rescaled with $\epsilon^{-1}$ since \eqref{eq:MMCase2} is for the concentration variable $x$ instead of $n$. We have
\begin{eqnarray*}
\lim_{\epsilon\to 0+} \epsilon\frac{k_2[S]}{[S]+k_M}
&=&\lim_{\epsilon\to 0+} \epsilon\frac{\tilde  k_1\epsilon^{-1} \tilde k_2\epsilon^{-1}}{\tilde k_1 \epsilon^{-1}+\tilde  k_{-1} \epsilon^{-1}+\tilde k_2 \epsilon^{-1}}=\frac{\tilde k_1\tilde k_2}{\tilde k_1+ \tilde k_{-1}+\tilde k_2}.
\end{eqnarray*}
Furthermore, the second order expansion of $H_x$ with respect to $p$
\begin{equation}
\frac{\partial ^2 H_x}{\partial p^2} \Big|_{p = 0} = \frac{2\tilde k_1^2\tilde k_2^2}{(\tilde k_1+\tilde k_{-1} + \tilde k_2)^3}+\frac{\tilde k_1\tilde k_2}{\tilde k_1+\tilde k_{-1} + \tilde k_2}+ \gamma x
\end{equation}
yields the following chemical Langevin approximation
\begin{eqnarray}
dx&=&\left(\frac{\tilde k_1\tilde k_2}{\tilde k_1+\tilde k_{-1} + \tilde k_2}-\gamma x\right)dt+\sqrt{\epsilon}\left(\sqrt{\frac{\tilde k_1\tilde k_2}{\tilde k_1+\tilde k_{-1} + \tilde k_2}+\frac{2\tilde k_1^2\tilde k_2^2}{(\tilde k_1+\tilde k_{-1} + \tilde k_2)^3}}dw_1-\sqrt{\gamma x}dw_2\right).
\end{eqnarray}

Although the above result is quite natural based on our derivations in previous sections, the application to Michaelis-Menten system again tells us that the strict correspondence between the drift and diffusion terms in classical large volume limit is lost.

\vspace*{2mm}
 {\bf (ii). Case 1:} $\alpha>1$.

 \vspace*{2mm}
 In this regime, $k_1[S]$ and $k_{-1}$ are much larger than $k_2$.  Similar as in Case 2, we have the transition probability density \eqref{eq:TransitionPDF} with Lagrangian \eqref{eq:MML}, and thus the Hamiltonian \eqref{eq:MMH} for the slow variable $x$. We get
\begin{eqnarray}\notag
 H_x(x,p)&=&\mathop{\sup}_{\cb}\mathop{\inf}_\varphi\left\{\tilde k_2 c_2(e^p-1)+\gamma x(e^{-p}-1)+c_2\left(\frac{\tilde k_{-1}}{\epsilon^{\alpha-1}}+\tilde k_2 e^p\right)(e^\varphi-1)+c_1\frac{\tilde k_{1}}{\epsilon^{\alpha-1}}(e^{-\varphi}-1)\right\}\\
 &=&\mathop{\sup}_{\cb}\left\{\tilde k_2c_2(e^p-1)+\gamma x(e^{-p}-1)-\frac{1}{\epsilon^{\alpha-1}}\left(\sqrt {c_2\left(\tilde k_{-1}+\tilde k_2e^p\epsilon^{\alpha-1}\right)}-\sqrt{c_1\tilde k_{1}}\right)^2\right\}.
\end{eqnarray}
As $\epsilon \to 0+$, the singular perturbation analysis suggests the term involving $1/\epsilon^{\alpha-1}$ to be 0, which gives $c_2=\tilde k_1/(\tilde k_{-1}+\tilde k_1)$. We obtain
\begin{eqnarray}\label{eq:MMCase1H}
 H_x(x,p)=\frac{\tilde k_1\tilde k_2}{\tilde k_1+ \tilde k_{-1}}(e^p-1)+\gamma x(e^p-1).
\end{eqnarray}
The mean field limit
 \begin{equation}
\frac{dx}{dt}=\frac{\tilde k_1\tilde k_2}{\tilde k_1+ \tilde k_{-1}}-\gamma x.
\end{equation}
Its consistency with the Michaelis-Menten law is straightforward by checking
\begin{eqnarray*}
\lim_{\epsilon\to 0+} \epsilon\frac{k_2[S]}{[S]+k_M}
&=&\lim_{\epsilon\to 0+} \epsilon\frac{\tilde  k_1\epsilon^{-\alpha} \tilde k_2\epsilon^{-1}}{\tilde k_1 \epsilon^{-\alpha}+\tilde  k_{-1} \epsilon^{-\alpha}+\tilde k_2 \epsilon^{-1}}=\frac{\tilde k_1\tilde k_2}{\tilde k_1+ \tilde k_{-1}}
\end{eqnarray*}
for $\alpha>1$.

In this regime the chemical Langevin approximation takes the form
\begin{eqnarray}
dx&=&\left(\frac{\tilde k_1\tilde k_2}{\tilde k_1+ \tilde k_{-1}}-\gamma x\right)dt+\sqrt{\epsilon}\left(\sqrt{\frac{\tilde k_1\tilde k_2}{\tilde k_1+ \tilde k_{-1}}}dw_1-\sqrt{\gamma x}dw_2\right)
\end{eqnarray}
and we formally recover the correspondence between the drift and diffusion terms in this case.

\vspace*{2mm}

 {\bf (iii). Case 3:} $0<\alpha<1$.

\vspace*{2mm}
In this regime, we need to select the scaling parameter for $n$ as $\epsilon^\alpha$, namely  to define $x_\alpha = n\epsilon^\alpha$. We have
\begin{equation}
 P(n_f,\tau | n_i,0) \propto \int Dx_\alpha D\cb \exp \left(-\frac{1}{\epsilon^\alpha}\int_{0}^{\tau}  L(x_\alpha,\dot x_\alpha,\cb) dt\right),
\end{equation}
where
 \begin{eqnarray}
L(x_\alpha,\dot x_\alpha,\cb)&=\mathop{\sup}_p\mathop{\sup}_\varphi&\left\{p \dot x_\alpha-\frac{\tilde k_2}{\epsilon^{1-\alpha}} c_2 (e^p-1)-\gamma x_\alpha(e^{-p}-1)\right.\notag\\
&&\left.-c_2\left(\tilde k_{-1}+\frac{\tilde k_2 e^p}{ \epsilon^{1-\alpha}}\right)(e^\varphi-1)-c_1\tilde k_{1}(e^{-\varphi}-1)\right\}.
\end{eqnarray}
The Hamiltonian corresponding to variable $x_\alpha$ has the form
\begin{eqnarray*}
 H_x(x_\alpha,p)&=&\mathop{\sup}_{\cb}\mathop{\inf}_\varphi\left\{\frac{\tilde k_2}{\epsilon^{1-\alpha}} c_2 (e^p-1)+\gamma x_\alpha(e^{-p}-1)+c_2\left(\tilde k_{-1}+\frac{\tilde k_2 e^p}{ \epsilon^{1-\alpha}}\right)(e^\varphi-1)+c_1\tilde k_{1}(e^{-\varphi}-1)\right\}\notag\\
 &=&\mathop{\sup}_{\cb}\left\{\frac{\tilde k_2}{\epsilon^{1-\alpha}}c_2(e^p-1)+\gamma x_\alpha(e^{-p}-1)-\left(\sqrt {c_2\left(\tilde k_{-1}+\frac{\tilde k_2e^p}{\epsilon^{1-\alpha}}\right)}-\sqrt{c_1\tilde k_{1}}\right)^2\right\}.
\end{eqnarray*}
As $\epsilon \to 0+$, the singular perturbation analysis gives $c_2=\epsilon^{1-\alpha}\tilde k_1e^p/\tilde k_2$, and the final Hamiltonian
\begin{eqnarray}\label{eq:MMCase3H}
 H_x(x_\alpha,p)=\tilde k_1(e^p-1)+\gamma x_\alpha(e^{-p}-1).
\end{eqnarray}
We get the mean field ODE
\begin{equation}
\frac{dx_\alpha}{dt}=\tilde k_1-\gamma x_\alpha,
\end{equation}
The consistency with the Michaelis-Menten law can be verified by checking
$$
\lim_{\epsilon\to 0+} \epsilon^\alpha\frac{k_2[S]}{[S]+k_M}
=\lim_{\epsilon\to 0+} \epsilon^\alpha\frac{\tilde  k_1\epsilon^{-\alpha} \tilde k_2\epsilon^{-1}}{\tilde k_1 \epsilon^{-\alpha}+\tilde  k_{-1} \epsilon^{-\alpha}+\tilde k_2 \epsilon^{-1}}=\tilde k_1
$$
for $0<\alpha<1$. In this regime we have the chemical Langevin approximation as
\begin{eqnarray}\label{eq:MMLangevinCase3}
dx_\alpha&=&\left(\tilde k_1-\gamma x_\alpha\right)dt+\sqrt{\epsilon^\alpha}\left(\sqrt{\tilde k_1}dw_1-\sqrt{\gamma x_\alpha}dw_2\right).
\end{eqnarray}
Again the formal correspondence between the drift and diffusion terms is recovered but with special concentration variable definition and rescaling. It is instructive to compare \eqref{eq:MMLangevinCase3} and \eqref{Langevin approximation 2} in the case $0<\alpha<1$. We have an additional term $\sqrt{\gamma x_\alpha}dw_2$ in \eqref{eq:MMLangevinCase3} because  of the utilized scaling $x_\alpha=n\epsilon^\alpha$ instead of $x=n\epsilon$ in the two-state model.
This reveals the difference between the single-molecule Michaelis-Menten and the two-state genetic switching model. Indeed in this regime, the reactions can be simplified  to
$$
\text{S}  \autorightarrow{$ k_1$}{} \text{P} \autorightarrow{$\gamma$}{}  \emptyset
$$
since the production rate is limited by the rate of formation of complex $ES$.

We summarize our findings for the single-molecule Michaelis-Menten kinetics in three regimes in Table \ref{Tab:Tab2}.
\renewcommand{\arraystretch}{2}
\begin{table}[htbp]
\centering
\caption{\com{The LDTs, mean-field limits and Langevin approximations for
single-molecule Michaelis-Menten kinetics}}\label{Tab:Tab2}
\begin{tabularx}{0.93\textwidth}{c |c|c|c}
\Xhline{1.2pt}\hline
    &\com{ LDT Hamiltonian} & \com{Deterministic Drift}  & \com{Noise in Langevin Approximation}  \\ \Xhline{1.2pt}
    {\com{\bf Case 1:}} \com{($\alpha>1$)} & \com{$H_x(x,p)$ in \eqref{eq:MMCase1H}} & \com{$\displaystyle \frac{ \tilde k_1\tilde k_2}{\tilde k_1+\tilde k_{-1}}-\gamma x$} & \Gape[1.2mm]{\com{$\displaystyle\sqrt{\epsilon}\left(\sqrt{\frac{\tilde k_1\tilde k_2}{\tilde k_1+ \tilde k_{-1}}}dw_1-\sqrt{\gamma x}dw_2\right)$}} \\ \hline
    {\com{\bf Case 2:} }\com{($\alpha=1$) } & \com{$H_x(x,p)$ in \eqref{eq:MMCase2H}}  &  \com{$\displaystyle\frac{\tilde k_1\tilde k_2}{\tilde k_1+\tilde k_{-1} + \tilde k_2}-\gamma x$}      &  \Gape[1.2mm]{\com{$\displaystyle\sqrt{\epsilon}\left(\sqrt{\frac{\tilde k_1\tilde k_2}{\tilde k_1+\tilde k_{-1} + \tilde k_2}+\frac{2\tilde k_1^2\tilde k_2^2}{(\tilde k_1+\tilde k_{-1} + \tilde k_2)^3}}dw_1-\sqrt{\gamma x}dw_2\right)$}} \\ \hline
    {\com{\bf Case 3:}} \com{ ($\alpha<1$) } & \com{$H_x(x,p)$ in \eqref{eq:MMCase3H}} &  \com{$\tilde k_1-\gamma x_\alpha$} & \Gape[1.2mm]{\com{$\displaystyle\sqrt{\epsilon^\alpha}\left(\sqrt{\tilde k_1}dw_1-\sqrt{\gamma x_\alpha}dw_2\right)$ }}  \\ \Xhline{1.2pt}
    \end{tabularx}%
\end{table}%
}

\section{Discussions and Conclusion}\label{sec:disc}

\begin{figure}[H]
\centering
\includegraphics[width=0.5\textwidth]{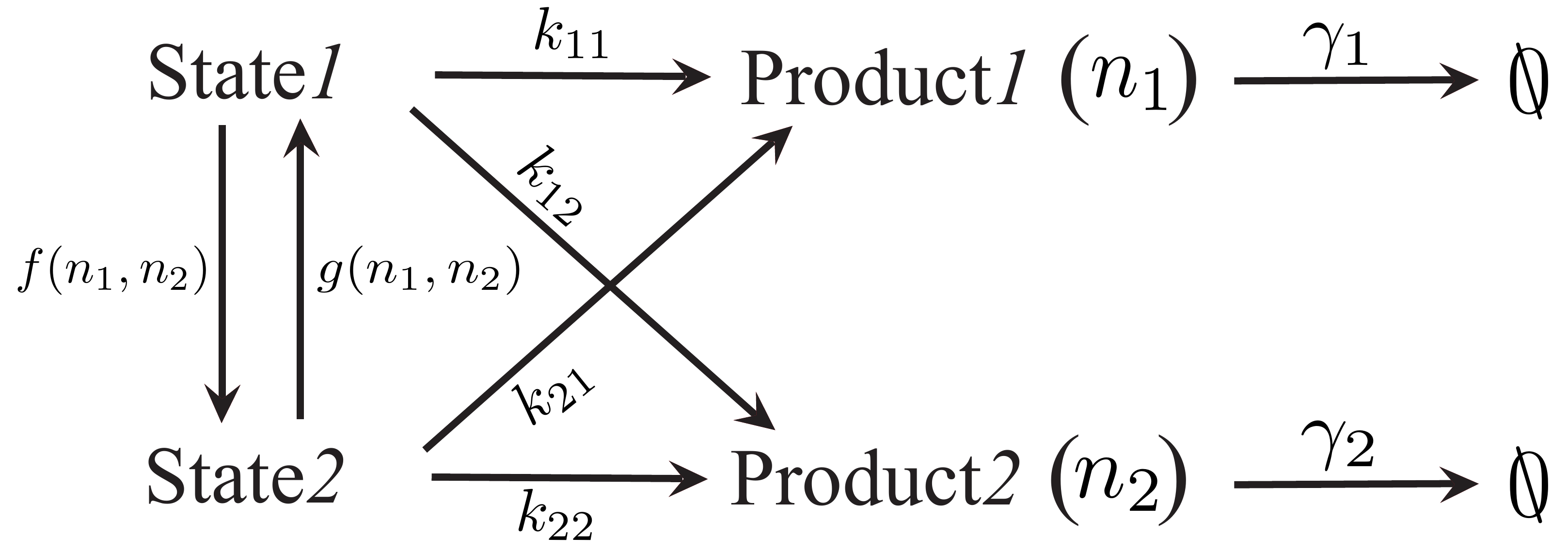}
\caption{Schematics of a two-scale kinetic model with two kinds of products.}\label{fig:2}
\end{figure}

The methods and LDT results we proposed in this paper are not limited to the two-state model, single-molecule Michaelis-Menten and single kind of product case. It is indeed general for a class of two-scale kinetic systems. To show this, let us consider the following extension as shown in Fig. \ref{fig:2}.

We assume similar scaling as considered in \eqref{eq:scaling}:
\begin{eqnarray*}
k_{ij}\sim \frac{1}{\epsilon},\quad f, g \sim \frac{1}{\epsilon^\alpha}, \quad \alpha>0,\quad (i,j=1,2).
\end{eqnarray*}
Define $x_j=n_j\epsilon$, $\tilde k_{ij}=k_{ij}\epsilon$ for $i,j=1,2$ and $\tilde f(x_1,x_2) = f(n_1,n_2)\epsilon^\alpha$, $\tilde g(x_1,x_2) = g(n_1,n_2)\epsilon^\alpha$.
Performing the same approach as in Sec. \ref{sec:two}, we get the transition probability
\begin{equation*}
 P(n_f,\tau | n_i,0) \propto \int D\xb  D\cb \exp \left(-\frac{1}{\epsilon}\int dt L_1(\xb,\dot \xb,\cb)-\frac{1}{\epsilon^\alpha}\int dt L_2(\xb,\cb)\right).
\end{equation*}
Here the Lagrangian
\begin{equation*}
L_1(\xb,\dot {\xb},\cb)=\mathop{\sup}_{\pb}\{\pb \cdot \dot {\xb}-\tilde H_1(\xb,\cb,\pb)\},\quad  L_2(\xb, \cb)=\mathop{\sup}_{\varphib} \{-\tilde H_2(\xb,\cb,\varphib)\},
\end{equation*}
where $\xb=(x_1,x_2)$, $\cb=(c_1,c_2)$, $\pb=(p_1,p_2)$, $\varphib=(\varphi_1,\varphi_2)$ and
\begin{eqnarray*}
 \tilde H_1(\xb,\cb,\pb)&=& (\tilde k_{11}c_1+\tilde k_{21}c_2)(e^{p_1}-1)+\gamma_1 x_1 [e^{-p_1}-1]+(\tilde k_{12}c_1+\tilde k_{22}c_2)(e^{p_2}-1)+\gamma_2 x_2 (e^{-p_2}-1),\\
 \tilde H_2(\xb,\cb,\varphib)&=& c_1 \tilde g(x)(e^{{\varphi}_1-{\varphi}_2}-1) + c_2 \tilde f(x)(e^{{\varphi}_2-{\varphi}_1}-1).
\end{eqnarray*}
All of the analysis performed for the two-state model can be applied here to  obtain the LDTs for variable $\xb=(x_1,x_2)$ with different $\alpha$.

One can also employ the WKB ansatz $P_{j}(x)\sim\exp(-\epsilon^{-1}\Phi_{j}(x))$  for the stationary distribution of the stochastic hybrid system \eqref{eq:HybridSystem}, where $x=n \epsilon$ and $j$ is a state of DNA. In the asymptotics, one gets a static Hamilton-Jacobi equation for the quasi-potential $\Phi_{j}$ and it turns out $\Phi_{j}$ does not depend on the specific choice of $j$. \com{However if not handled appropriately, the WKB approximation may lead to totally different forms of Hamiltonian \cite{LiLin} as mentioned in the end of Section \ref{sec:example}.} This non-uniqueness is due to the lack of variational selection in LDT, which gives a unique Hamiltonian dual to the obtained Lagrangian in rate functional. And this Hamiltonian has the superiority that it is convex with respect to the momentum variable as the by-product of Legendre-Fenchel transform and LDT. This property is important for the nice behavior of numerical discretization.

In this paper, we assume the switching rates between different DNA states are in order $\epsilon^{-\alpha}$. It is not necessary and could be more general. As long as the switching rates between different DNA states are in $O(\lambda(\epsilon))$, and the cases $\lim_{\epsilon \to 0+} \epsilon\lambda(\epsilon)=\infty,~O(1)$ and $0$ are considered, we will get similar results.  Especially, the readers may easily verify that if we assume $\lambda(\epsilon)=K\epsilon^{-1}$, then the two-scale LDT Lagrangian with $\epsilon$-scaling in front of  $-\ln P$ has the form
\begin{eqnarray}
L_x(x,\dot x)=\mathop{\inf}_{\cb}\left\{L_1(x,\dot x,\cb)+KL_2(x,\cb)\right\},
\end{eqnarray}
and the LDT Lagrangian with $\lambda(\epsilon)$-scaling in front of  $-\ln P$ has the form
\begin{eqnarray}
L_x(x,\dot x)=\mathop{\inf}_{\cb}\left\{K^{-1}L_1(x,\dot x,\cb)+L_2(x,\cb)\right\}.
\end{eqnarray}
When $K$ goes to $0, 1$ or $\infty$, the appropriate choices of scaling recover the desired results shown in the paper.

In conclusion, we established  the two-scale LDTs for a class of chemical reaction kinetics through
the second-quantization path integral formulation. Although not rigorous, we showed that this formal approach is very effective and transparent to understand the two-scale LDTs associated with different reaction channels. This provides essential insights to rigorously prove the corresponding LDTs, which is our ongoing research. We discussed its implication on single-molecule Michaelis-Menten kinetics as well. The proposed framework and results also shed lights on the understanding of general multi-scale systems including diffusion processes. It will be interesting to investigate the application of two-scale LDTs to other systems.

\section*{Acknowledgement}

T. Li acknowledges the support of NSFC under grants 11171009, 11421101, 91130005 and the National Science Foundation for Excellent Young Scholars (Grant No. 11222114). They also thank Weinan E, Yong Liu, and Xiaoguang Li for helpful discussions.

\section*{References}

\bibliographystyle{}

\begin{thebibliography}{10}

\bibitem{Assaf}
Assaf M, Roberts E and Luthey-Schulten Z 2011
 Determining the Stability of Genetic Switches: Explicitly Accounting for mRNA Noise \newblock{\em Phys. Rev. Lett.} 106  2048102.

 \bibitem{Orszag}
Bender C M and  Orszag S A  1999
 \newblock{\em Advanced Mathematical Methods for Scientists and Engineers: Asymptotic Methods and Perturbation Theory} (New York: Springer) .

\bibitem{Bressloff}
Bressloff P C and Faugeras O 2014 On the Hamiltonian structure of large deviations in
stochastic hybrid systems \newblock{\em arXiv:1410.2152v1}.

\bibitem{Newby13}
 Bressloff P C and  Newby J M 2013
Metastability in a Stochastic Neural Network Modeled as a Velocity Jump Markov Process  \newblock{\em  SIAM Appl. Dyn. Syst.} 12  1394.

\bibitem{Buhler}
B\"{u}hler O  2006
\newblock{\em A Brief Introduction to Classical, Statistical and Quantum Mechanics} (Providence: American Mathematical Society).

\bibitem{Dembo}
Dembo A and Zeitouni O  1998 \newblock{\em Large deviations techniques and applications, 2nd edition} (New York: Springer-
Verlag).

\bibitem{Doi}
Doi M 1976
 Second quantization representation for classical many-particle system  \newblock{\em J. Phys. A: Math. Gen.} 9  1465.

 \bibitem{FAGGIONATO}
Faggionato A, Gabrielli D  and  Crivellari M R 2010  Averaging and large deviation principles for fully-coupled piecewise deterministic Markov processes and applications to molecular motors \newblock{\em Markov Process. Relat. Fields} 16  497.

\bibitem{Freidlin}
Freidlin M I and Wentzell A D 1998 \newblock{\em Random perturbations of dynamical systems, 2nd edition} (New York: Springer) .

\bibitem{Ge}
Ge H, Qian H  and Xie X S  2015
Stochastic phenotype transition of a single cell in an intermediate region of gene state switching \newblock{\em  Phys. Rev. Lett.} 114  078101.

\bibitem{gMAM}
Heymann M and Vanden-Eijnden E 2008  The geometric minimum action method: A least action principle on the space of curves \newblock{\em Comm. Pure Appl. Math.} 61  1052.

\bibitem{Keener}
Keener J and Sneyd J 1998
\newblock{\em Mathematical Physiology} (New York: Springer-Verlag).

\bibitem{Kemeny}
 Kemeny J G and Snell J L 1960
\newblock{\em Finite Markov chains} (New York, Berlin and Heidelberg: Springer-Verlag) .

\bibitem{Xie}
Kou S C, Cherayil B J, Min  W, English B P and Xie X S 2005  Single-molecule Michaelis-Menten Equations  \newblock{\em J. Phys. Chem. B} 109  19068.

\bibitem{LiLin}
Li T and Lin F 2015
Large deviations for two-scale chemical kinetic processes \newblock{\em arXiv:1504.03781}.

\bibitem{Liptser1996}
Liptser R 1996
 Large deviations for two scaled diffusions   \newblock{\em Prob. Theory Relat. Fields} 106  71.

\bibitem{Lv14}
Lv C , Li X,  Li F and  Li T 2014
 Constructing the energy landscape for genetic switching system driven by intrinsic noise \newblock{\em PLoS ONE} 9  e88167.

\bibitem{Lv15}
Lv C , Li X,  Li F and  Li T 2014
Energy landscape reveals that the budding yeast cell cycle is a robust and adaptive multi-stage process \newblock{\em PLoS Comp. Biol.} 9 e88167.

\bibitem{Michaelis Menten}
Michaelis L and  Menten M L 1913  Die Kinetik der Invertinwerkung \newblock{\em Biochem. Z.} 49  333.

\bibitem{Newby Phys Biol}
Newby J M 2012  Isolating intrinsic noise sources in a stochastic genetic switch \newblock{\em Phys. Biol.} 9 026002.

\bibitem{Newby Dyn Syst}
Newby J M 2014 Spontaneous Excitability in the Morris¨CLecar Model with Ion Channel Noise \newblock{\em  SIAM J. Appl. Dyn. Syst.} 13  1756.

\bibitem{Newby10}
 Newby J M and Bressloff P C 2010
  Local synaptic signaling enhances the stochastic transport of motor-driven cargo in neurons  \newblock{\em Phys. Biol.} 7  036004.

\bibitem{Newby Math Biol}
Newby J M and  Chapman J 2014  Metastable behavior in Markov processes with internal
states  \newblock{\em  J. Math. Biol.} 69  941.

\bibitem{Peliti}
Peliti L 1985
 Path integral approach to birth-death processes on a lattice \newblock{\em  J. Phys.} 46  1469.

\bibitem{Rockafellar}
  Rockafellar R T and Wets R J B 1998 \newblock{\em Variational Analysis} (Berlin and HeidelbergSpringer) .

\bibitem{Shwartz1995}
Shwartz A and Weiss A  1995
\newblock{\em Large deivations for performance analysis: queues, communications and computing} (London: Chapman and Hall).

\bibitem{Hugo}
Touchette H 2009
 The large deviation approach to statistical mechanics \newblock{\em Phys. Rep.} 478  1.

\bibitem{Varadhan}
Varadhan S R S 1984 \newblock{\em Large deviations and applications} (Philadelphia: SIAM)  .

\bibitem{Veretennikov2000}
Veretennikov A Yu 2000
 On large deviations for SDEs with small diffusion and averaging \newblock{\em Stoch. Process. Appl.} 89  69.

\bibitem{Veretennikov1999}
Veretennikov A Yu 1999
 On large deviations in the averaging principle for SDE's with a "full
dependence" \newblock{\em Ann. Prob.} 27  284.


\bibitem{Wolyness}
Zhang B and  Wolyness P G 2014
Stem cell differentiation as a many-body problem \newblock{\em  Proc. Natl. Acad. Sci. U.S.A.} 111  10185.

\bibitem{Wang}
Zhang K, Sasai M and Wang J 2013
Eddy current and coupled landscapes for nonadiabatic and nonequilibrium complex system dynamics \newblock{\em  Proc. Natl. Acad. Sci. U.S.A.} 110  14930.


\bibitem{ZhouLi}
Zhou P and Li T 2015
 Realization of Waddington's metaphor: Potential landscape, quasi-potential, A-type integral and beyond \newblock{\em  arXiv:1511.02088} .



\end{thebibliography}

\end{document}